\documentclass[12pt]{article}
\pdfoutput=1
\usepackage{geometry,enumerate,amsmath,amssymb}
\usepackage{fullpage}
\usepackage{graphicx}
\usepackage{float}
\newcommand{\be}{\begin{equation}}
\newcommand{\ee}{\end{equation}}

\newcommand{\news}{\setcounter{equation}{0}\quad}
\def\bea{\begin{eqnarray}}
\def\eea{\end{eqnarray}}
\begin{document}
\title{
\begin{flushright}\ \vskip -2cm {\small{\em DCPT-14/07}}\end{flushright}
Leapfrogging vortex rings\\ in the Landau-Lifshitz equation\\[10pt]}
\author{
Antti J. Niemi$^\star$ and 
Paul Sutcliffe$^\dagger$\\[10pt]
{\em \normalsize $^\star$
{Department of Physics and Astronomy},
{Uppsala University,}
}\\
{\em \normalsize Box 803, S-75108, Uppsala, Sweden,}\\
{\em \normalsize {Laboratoire de Mathematiques et Physique Theorique},
{CNRS UMR 6083,}}\\
{\em \normalsize
 F\'ed\'eration Denis Poisson,
{ Universit\'e de Tours},}\\
{\em \normalsize
{Parc de Grandmont, F37200, Tours, France, and}} \\
{\em \normalsize
{Department of Physics, Beijing Institute of Technology}},\\
 {\em \normalsize
{Haidian District, Beijing 100081, P. R. China}}\\
{\em \normalsize $^\dagger$ Department of Mathematical Sciences, Durham University,}\\
 {\em \normalsize
Durham DH1 3LE, U.K.}\\[0pt]
{\em \normalsize Email:\quad 
{Antti.Niemi@physics.uu.se}, \quad p.m.sutcliffe@durham.ac.uk}\\ 
}
\date{February 2014, version to appear in Nonlinearity}

\maketitle
\begin{abstract}
Vortex rings are ubiquitous in fluids, with smoke rings being a 
familiar example.
The interaction of multiple vortex rings produces complex dynamical 
behaviour, such as the leapfrogging motion 
first analysed by Helmholtz more than a century and a half ago.
Here we report on numerical investigations of vortex ring dynamics in a 
different setting from fluids, namely, as solutions of the 
Landau-Lifshitz 
equation that models the evolution of
the local magnetization in a ferromagnetic medium. 
We present the results of the first study on
the dynamics of interacting magnetic vortex rings
and provide a novel link between fluids and magnetism, by showing that a
range of phenomena familiar in fluids are reproduced in ferromagnets.
This includes the leapfrogging motion of a pair of vortex rings 
and evidence for the chaotic dynamics of a trio of rings.\\


\end{abstract}
\newpage
\section{Introduction}\news
Vortex rings in fluids are well-studied using a variety of
analytic, numerical and experimental techniques (for a review
see \cite{SL}).
The interaction of multiple vortex rings produces complex dynamical 
behaviour, such as the leapfrogging locomotion of a pair of
vortex rings.
The number of leapfrogs increases with 
Reynolds number \cite{RS} and perpetual leapfrogging 
can only occur in inviscid flow.
Numerical simulations of the Euler equations can reproduce
several leapfrogs, providing sophisticated methods are used
to deal with numerical diffusion \cite{SU}.
Studies using a thin-cored approximation accurately describe
leapfrogging and reveal that the dynamics of 
three or more coaxial vortex rings has 
regimes of chaotic behaviour in which the
evolution is very sensitive to the initial condition \cite{Ko}.
The interaction of two vortex rings in a superfluid shares many
common features with those in a normal fluid, as demonstrated by
numerical simulations of the Gross-Pitaevskii equation \cite{KL},
so some phenomena may be universal for all types of vortex rings.

The Landau-Lifshitz equation describes the dynamics  
of the local magnetization in a ferromagnetic medium.
This equation has vortex ring solutions
that are magnetic analogues of vortex rings in fluids.
They are toroidal regions in which the 
magnetization is not in the ground state,
 and they propagate along their symmetry axis with a 
constant speed \cite{Pa,Co,Su}.
The existence of vortex ring solutions of the Landau-Lifshitz equation 
is a fairly recent result, and so far studies have been limited.
In fact, all previous work has been restricted to the case of a single
magnetic vortex ring in uniform motion, so the full extent to which
these are indeed analogues of vortex rings in fluids remains to be seen.
Investigating this issue is the main aim of the current paper.

Here we present the first results on the dynamics and interaction of multiple 
magnetic vortex rings, obtained from numerical simulations of the 
Landau-Lifshitz equation. In particular, we demonstrate the
leapfrogging motion of a pair of magnetic vortex rings and 
evidence for the chaotic dynamics of a trio of rings. 
The Landau-Lifshitz equation is more amenable to a standard
numerical treatment than the Euler or Navier-Stokes equations.
As a result, we are able to apply a simple finite difference
scheme with a Runge-Kutta algorithm to 
compute the evolution of several coaxial rings.
The direct simulation of the nonlinear partial differential equation
removes the need to resort to any form of thin-cored approximation.
Within the thin-cored approximation of fluid vortex rings,
Shashikanth and Marsden \cite{SM} have shown that 
the periodic leapfrogging motion of a pair of  
rings generates a geometric phase. 
These authors state that they view this work as a preliminary step to more
sophisticated modelling of ring interactions using ideas from geometric 
mechanics. Our results show that the Landau-Lifshitz equation is an ideal 
system in which to apply these ideas, beyond the  
thin-cored approximation, to a tractable partial differential equation.
Our demonstration of periodic leapfrogging of vortex rings indicates
that this system could be useful for any future investigations in this 
direction.

Our numerical simulations of the Landau-Lifshitz equation suggests
a novel link between fluids and magnetism, with many familiar
phenomena for fluid vortex rings being reproduced in the nanoscale
world of ferromagnetic media. The interesting properties of
magnetic vortex rings that we have demonstrated provides some motivation 
for attempts at experimental creation and observation of these 
structures. If this can be achieved then there are potential applications
within the field of spintronics, and we shall speculate on this possibility
later in this paper.

The outline of the paper is as follows. In section \ref{sec-LL}
we review the Landau-Lifshitz equation and describe its vortex ring
solution. 
In section \ref{sec-leapfrog} we present our main results on the 
dynamics of multiple vortex rings.
In section \ref{sec-LLG} we investigate the effect of including dissipation,
by solving the Landau-Lifshitz-Gilbert equation, and compare with
known results on the leapfrogging of fluid vortex rings for 
varying Reynolds number. Finally, we present some conclusions
in section \ref{sec-con}.

\section{A vortex ring in the Landau-Lifshitz equation}\news\label{sec-LL}
The system of interest is a three-dimensional ferromagnet with 
isotropic exchange interactions and an easy axis anisotropy. 
In the continuum approximation, the ferromagnet is
described by its spin ${\bf n}({\bf x},t),$
which is a three-component unit vector 
${\bf n}=(n_1,n_2,n_3),$ 
specifying the local orientation of the magnetization. 
In the absence of dissipation, the dynamics of the ferromagnet is
governed by the Landau-Lifshitz equation
\be
\frac{\partial {\bf n}}{\partial t}={\bf n} \times 
\{\nabla^2{\bf n} + A({\bf n}\cdot{\bf k}){\bf k}\}.
\label{ll}
\ee
Here ${\bf k}=(0,0,1)$ is the easy axis and we work in the unbounded
domain $\mathbb{R}^3,$ with the ground state aligned along the easy axis,
so that the boundary conditions are ${\bf n}\rightarrow {\bf k}$
as $|{\bf x}|\rightarrow \infty.$

For simplicity of presentation, we have used suitably scaled dimensionless
units. The corresponding physical units depend on the material properties, 
such as the effective spin exchange interaction.
For typical material parameters the dimensionless unit of length equates to 
a physical size of around 10 nanometres and the dimensionless unit of time to 
around 10 picoseconds (see \cite{TKS} for details). 
However, it is important to note that, as 
discussed below, the radius of a magnetic vortex ring is a 
parameter that can vary over a large range and is determined by the initial
conditions. In particular, the size of a vortex ring is not fixed by any 
constants in the Landau-Lifshitz equation, which are related to the 
material parameters.
This is, of course, entirely expected and mirrors the situation for 
fluid vortex rings.  As we shall see, both the size and speed of a ring 
can be adjusted by several orders of magnitude, 
which should make their physical existence 
realistic over a wide range of material conditions. The examples we present in
detail are chosen so that characteristic scales and speeds are of the same 
order of magnitude as for current-driven domain wall motion realized in 
magnetic wires \cite{Ya}, where typical speeds are of the order of 10m/s.

There is a scaling transformation \cite{Co} that relates magnetic vortex 
rings with 
differing values of $A$, so for computational concreteness we set 
$A=1$ from now on.
Numerical studies in \cite{Co} restrict to the isotropic case of $A=0$, 
but the investigations in \cite{Su} reveal that the results for axial 
vortex rings are 
similar for $A=0$ and $A=1$. Thus the value of $A$
 is not crucial for the phenomena 
described in this paper and we expect little qualitative difference between
hard materials and permalloy.

A magnetic vortex ring is an axially symmetric configuration 
that propagates at constant speed along its symmetry axis.
Far from the vortex ring the spin is in the ground state ${\bf n}={\bf k},$
whereas in the core of the ring the spin points in the antipodal direction 
${\bf n}=-{\bf k}.$ A useful way to visualize a magnetic vortex ring is
to plot the isosurface $n_3=0,$
delineating the boundary between the core where $n_3=-1$ and the vacuum 
where $n_3=1$.
Such an isosurface indicates both the position of 
the core of the ring and its thickness.  

The existence of a magnetic vortex ring 
as a stationary structure was first suggested in \cite{DI}, 
but it took a careful consideration of the 
conserved quantities of the Landau-Lifshitz equation to
identify that such structures could not be stationary, 
but instead must propagate at 
constant speed along their symmetry axis \cite{Pa}. 
Subsequently, an analysis \cite{Co} based on an axially symmetric 
travelling wave 
ansatz revealed that a magnetic vortex ring is characterized by two
real parameters. 
Roughly speaking, these parameters determine the 
radius and the thickness of the vortex ring, and 
the ring exists providing its radius is 
above a critical value determined by its thickness. 
In this paper we shall be concerned with vortex rings where the
radius of the ring is large in comparison to its thickness. 
As described in \cite{Su},
in this
situation the cross-section of a vortex ring at any given time 
is well-approximated
by a stationary solution (called a Skyrmion) 
of the two-dimensional version of the 
Landau-Lifshitz equation (\ref{ll}).
It is therefore useful to
briefly review some results on these planar Skyrmions.
For more details see \cite{PZ}.
It is perhaps worth pointing out that, despite the name 
magnetic vortex ring, its cross-section is a planar
Skyrmion and not a magnetic spin vortex \cite{Hu}.
Magnetic spin vortices appear in two-dimensional systems with an easy plane 
anisotropy, rather than the easy axis anisotropy considered here.

To obtain the planar Landau-Lifshitz equation from (\ref{ll}) we simply take 
the spin ${\bf n}$ to be independent of the third Cartesian coordinate $z.$
Introducing polar coordinates $(r,\phi)$ in the $(x,y)$ plane, a stationary
planar Skyrmion located at the origin has the form
\be
{\bf n}=(\sin f\cos(q\phi+\omega t),\sin f \sin(q\phi+\omega t), \cos f),
\label{ansatz}
\ee
where $\omega$ is the constant frequency of precession, $q$ is a non-zero
integer and 
$f(r)$ is a radial profile function satisfying the boundary conditions
 $f(0)=\pi$ and $f(r)\rightarrow 0$ as $r\rightarrow\infty.$

The Skyrmion is an example of a topological soliton \cite{MS} 
and the integer $q$ counts the
number of times that the unit vector ${\bf n}$ covers the two-sphere
as $(x,y)$ varies over the entire plane.
It is this kind of topological arrangement of the spin that is called a 
Skyrmion, or
sometimes a baby Skyrmion to emphasize its planar character.
The integer $q$ is known as the Skyrmion number, with $q<0$ called 
an anti-Skyrmion. Similar Skyrmions have recently been observed experimentally
in chiral ferromagnets \cite{Yu}, whose mathematical description
involves an additional term in the Landau-Lifshitz equation
that derives from a contribution to the energy 
(called the Dzyaloshinskii-Moriya interaction) 
that is first order in the derivative of the spin.
This extra term allows 
static Skyrmion solutions, in contrast to the
dynamical Skyrmions considered here that are stationary
but not static. However, the spatial structure
of the spin has the same form for Skyrmions in both systems.

Substituting the ansatz (\ref{ansatz}) into the Landau-Lifshitz equation 
(\ref{ll}) yields a stationary solution providing the profile function
satisfies the ordinary differential equation 
\be
f''+\frac{f'}{r}-\bigg(1+\frac{q^2}{r^2}\bigg)\sin f\cos f+\omega \sin f=0.
\label{ode}
\ee
\begin{figure}[h]\begin{center}
\includegraphics[width=8cm]{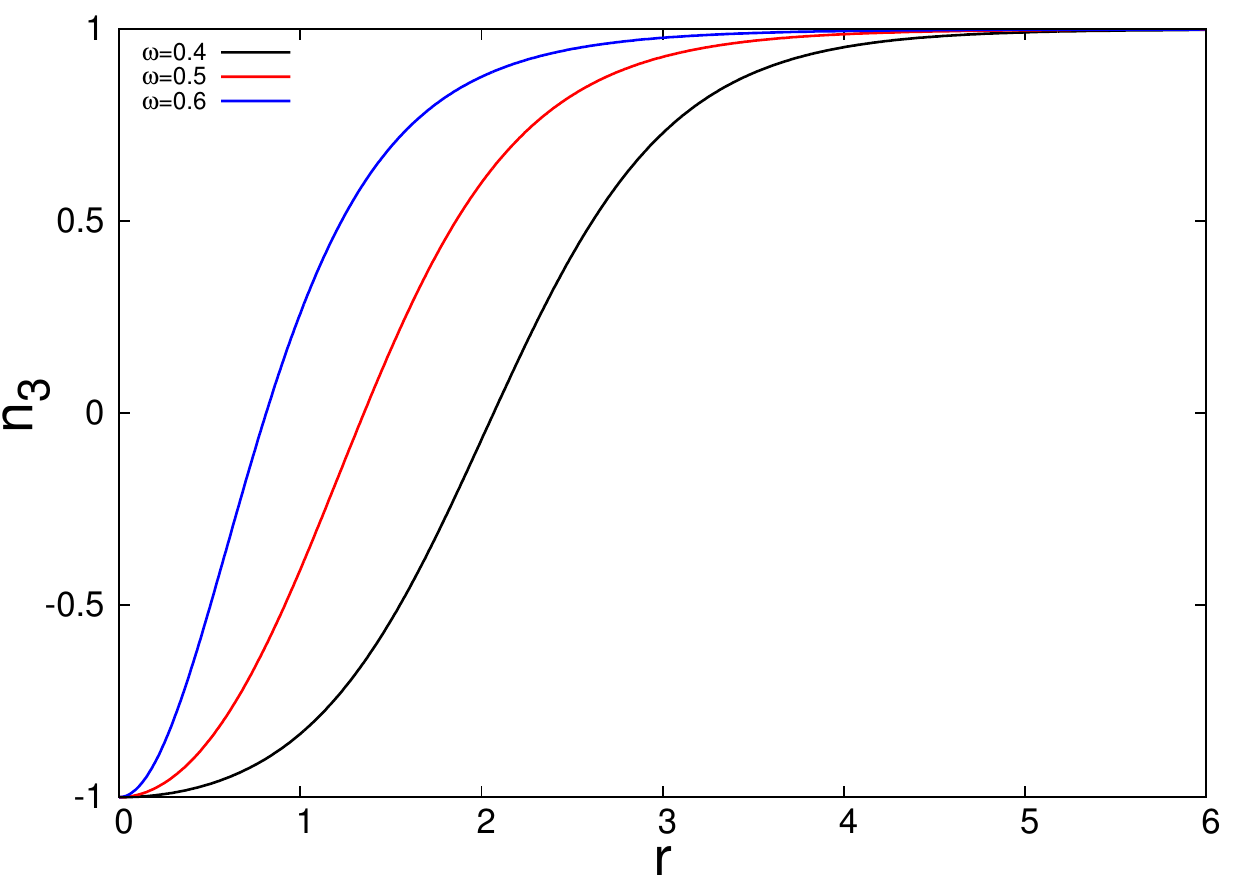}
\caption{
A plot of $n_3=\cos f$ against $r$ for planar Skyrmions with
frequencies $\omega=0.4$ (black curve);
$\omega=0.5$ (red curve);
$\omega=0.6$ (blue curve).
}
\label{fig-profiles}\end{center}\end{figure}
For a solution of this equation to exist, that satisfies the given
boundary conditions, the frequency must be restricted to the
range $0<\omega<1.$ More generally, the restriction is to
the range $0<\omega<A.$
Here we shall restrict attention to the simplest case of the single 
Skyrmion and set $q=1$ from now on. 
In terms of vortex rings, a
Skyrmion in the ferromagnetic system will play the role of
a vortex in the analogous fluid setting.

The frequency $\omega$ is a parameter of the Skyrmion, with
the size of the Skyrmion decreasing as the frequency increases.
As an illustration of this phenomenon, in Figure~\ref{fig-profiles}
we plot $n_3=\cos f$ as a function of $r$ for three increasing values
of the frequency $\omega=0.4,0.5,0.6.$ The position of the Skyrmion
is the point in space where ${\bf n}=-{\bf k},$ that is, the spin is
antipodal to the ground state spin. In the ansatz (\ref{ansatz}) this
position has been chosen to be the origin, but obviously the translation
invariance of the Landau-Lifshitz equation allows the Skyrmion to be
positioned at any point in the plane. As $n_3$ varies from the value $-1$
at the centre of the Skyrmion to the value $1$ at spatial infinity then
a natural definition for the size of the Skyrmion is the distance
from the centre at which $n_3$ vanishes. It is this definition that is
used in the above statement regarding the size of the Skyrmion decreasing
as $\omega$ increases. 

As mentioned briefly above, axially symmetric initial conditions for a
vortex ring can be constructed by embedding a Skyrmion along
a circle, so that a cross-section of the vortex ring is the 
Skyrmion \cite{Su}. 
Explicitly, introduce cylindrical coordinates $\rho,\theta,z$
where $x=\rho\cos\theta$ and $y=\rho\sin\theta,$ and impose axial
symmetry by requiring ${\bf n}$ to be independent of $\theta.$
The initial condition for a vortex ring of radius $R$ and position
$z_0$ along the symmetry axis is obtained by embedding the 
above planar Skyrmion (with frequency $\omega$) in the $(\rho,z)$ plane
with position $(\rho,z)=(R,z_0).$ The vortex ring is obtained by rotating
this planar configuration around the $z$-axis. 
The size of the Skyrmion (determined by $\omega$) 
corresponds to the thickness
of the vortex ring and the vortex ring radius is equal to $R.$
In terms of an explicit formula, this initial condition is given by
\be
{\bf n}=\bigg(\frac{(\rho-R)}{D}\sin F,\frac{(z-z_0)}{D}\sin F,\cos F\bigg),
\quad \mbox{where} \quad D=\sqrt{(\rho-R)^2+(z-z_0)^2}
\ee
and $F=0$ if $D>R$ but $F=f(D)$ if $D\le R,$ where 
$f(r)$ is the solution of the ordinary differential equation 
(\ref{ode}) with $q=1$ and the boundary
conditions $f(0)=\pi$ and $f(R)=0.$

By scaling symmetry, the important quantity for a vortex ring
is the ratio of its radius to its thickness, which we want to be large
for our investigations. Without loss of generality, we may therefore fix 
the size of the Skyrmion (that is, the thickness of the ring) and treat
the vortex ring radius $R$ as a parameter to vary. From now on,
we fix the frequency of the Skyrmion to be $\omega=0.5,$ which 
corresponds to a size of $1.35,$ as confirmed by an examination of
Figure~\ref{fig-profiles}. The regime in which we are interested is therefore
$R\gg 1.$

We study the dynamics of coaxial vortex rings 
by computing axially symmetric solutions of the Landau-Lifshitz 
equation (\ref{ll}). In cylindrical coordinates 
the  Landau-Lifshitz equation
reduces to the dynamics of an effective two-dimensional problem in the variables
$\rho$ and $z,$ because ${\bf n}$ is independent of $\theta.$
Spatial derivatives are approximated using
second order accurate finite differences with a lattice spacing
$\Delta \rho=\Delta z=0.15.$ 
The boundary condition on the symmetry axis is $\partial_\rho{\bf n}={\bf 0}$
and on the remaining boundaries of the numerical lattice we set 
${\bf n}={\bf k},$ so that the vacuum value is attained.  
Time evolution is implemented via a fourth order Runge-Kutta  
method with a timestep $\Delta t=0.006.$ 

\begin{figure}[ht]\begin{center}
\includegraphics[width=6cm]{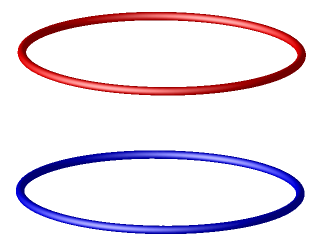}
\caption{
The isosurface $n_3=0$ at the initial time $t=0$ (lower blue ring)
and the later time $t=1800$ (upper red ring).
The vortex ring has a radius $R=45$ and moves with a constant
 speed $v=0.026.$ Between the two images it has travelled a 
distance $46.8$ which is comparable to its radius.
}
\label{fig-onering}\end{center}\end{figure}

\begin{figure}[ht]\begin{center}
\includegraphics[width=8cm]{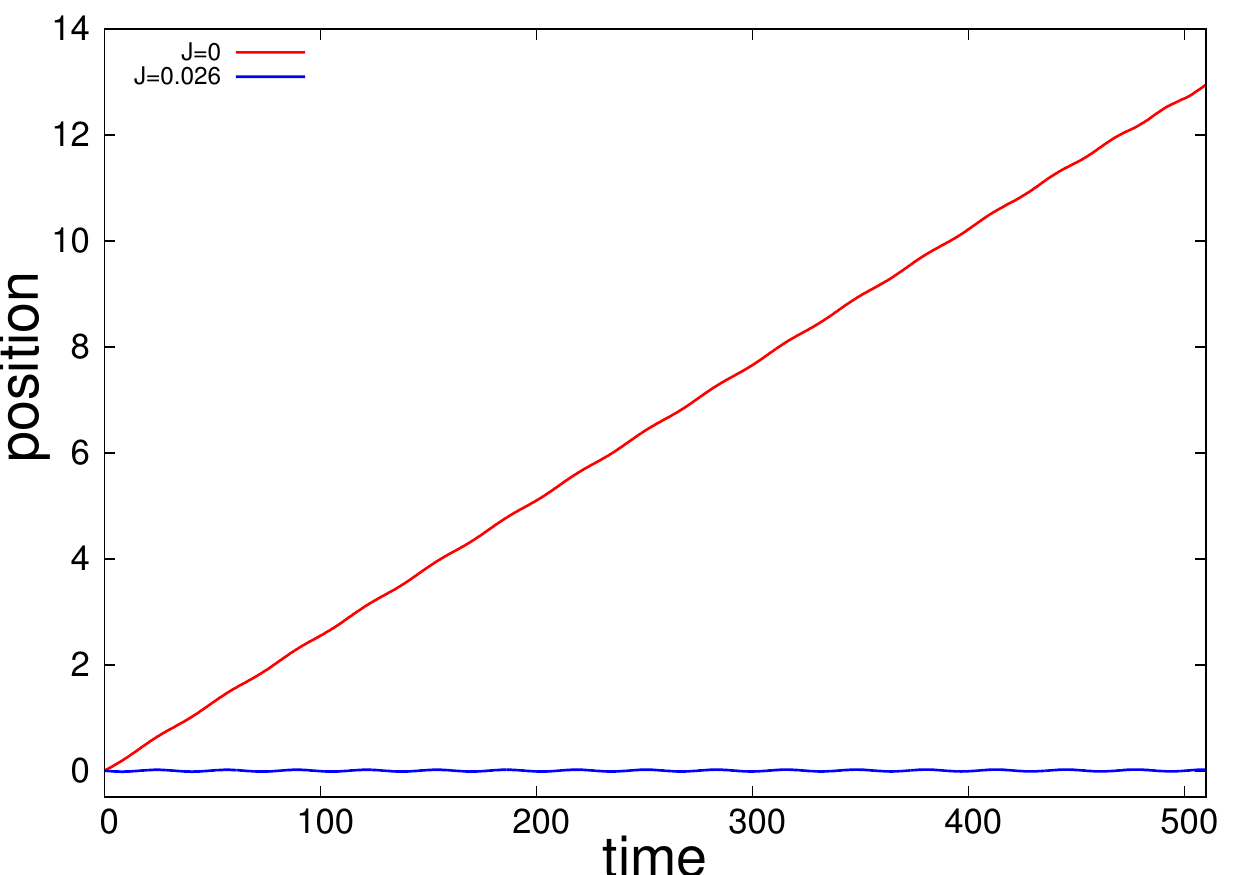}
\caption{
The (upper) red curve shows the position 
along the $z$-axis as a function of time 
for a vortex ring with a radius $R=45.$
The (lower) blue curve shows the position of the
same vortex ring when there is an applied electric current 
$J=0.026$ to freeze the motion.
}
\label{fig-position_J}\end{center}\end{figure}

In Figure~\ref{fig-onering} the (lower) blue ring is the isosurface
$n_3=0$ of the initial condition ($t=0$) obtained by embedding the 
Skyrmion (with $\omega=0.5$) along a circle of radius $R=45.$
The region for this numerical simulation is 
$(\rho,z)\in[0,90]\times[-45,45].$ 
The (upper) red ring in Figure~\ref{fig-onering}
is the $n_3=0$ isosurface at the later time
$t=1800.$ It can be seen that the vortex ring simply translates
along its symmetry axis with no change in the radius or thickness
of the ring. In Figure~\ref{fig-position_J} the (upper) red curve
shows the position along the $z$-axis as a function of time.
This confirms that the vortex ring travels at a constant speed, to
within a good numerical accuracy. The tiny deviations from
uniform motion are a result of using an initial condition that assumes
the cross-section is exactly that of a Skyrmion, 
rather than solving numerically for the initial conditions using a 
travelling wave ansatz.
This approximation is already accurate enough for our requirements when
the ring radius is $R=45,$ and the accuracy of this approximation
improves as the radius $R$ increases.

From the data in Figure~\ref{fig-position_J} we determine that
the speed of the ring is $v=0.026.$ 
From our earlier comment that the dimensionless unit of length equates to 
a physical size of around 10 nanometres and the dimensionless unit of time to 
around 10 picoseconds, then in terms of physical units this is a speed of 
the order of $10\,m/s,$
which is comparable to the speeds observed for  
current-driven domain wall motion in magnetic wires \cite{Ya}.
Once again, we stress that the speed of a vortex ring is not fixed by the
material parameters but by the size of the vortex ring. 
In particular, larger rings of the same thickness travel more slowly. 

In the presence of a spin-polarized electric current, ${\bf J},$
the Landau-Lifshitz equation (\ref{ll}) contains an extra term
and is given by \cite{TKS}
\be
\frac{\partial {\bf n}}{\partial t}={\bf n} \times 
\{\nabla^2{\bf n} + A({\bf n}\cdot{\bf k}){\bf k}\}
+({\bf J\cdot \nabla}){\bf n},
\label{llj}
\ee
where we continue to use dimensionless units.
An important feature is that the Landau-Lifshitz equation (\ref{llj}) 
has a Galilean symmetry
${\bf x}\mapsto {\bf x}-{\bf v}t,$ if accompanied by a corresponding
shift in the spin-polarized electric current 
${\bf J}\mapsto {\bf J}+{\bf v}.$
The microscopic derivation of this result 
and its physical interpretation is subtle. 
We refer the reader to section 14 of \cite{TKS}, and references
therein, for a detailed discussion.
The Galilean symmetry implies that a single vortex ring is brought to rest in 
the presence of an appropriate current ${\bf J}=(0,0,J),$
where $J$ is the speed of the vortex ring in the absence of an
electric current. This is a very useful feature 
that can be exploited in the numerical study of 
vortex rings.

Without an electric current, 
extremely large grids would be required to prevent the
rings from leaving the simulation region during the long 
time periods needed for the study of interacting vortex rings.
However, with a suitable electric current the centre of mass of the
system can be fixed, allowing the rings to remain inside a reasonable
simulation region for a long time.
As an example, we have seen that a vortex ring with a radius $R=45$
moves along the positive $z$-axis with a speed $v=0.026,$ 
when the current vanishes ${ J}={0}$. We can therefore bring this
vortex ring to rest by applying the current $J=v=0.026.$ 
The (lower) blue curve in 
Figure~\ref{fig-position_J} shows the position along the $z$-axis 
of the vortex ring as a function of time, obtained by numerical simulation of 
equation (\ref{llj}) with this value of the current.
This confirms that the introduction of this specific current freezes
the motion of the centre of mass of the ring. 

\section{Interacting vortex rings}\news\label{sec-leapfrog}
In this section we investigate the interaction of multiple vortex
rings and demonstrate that leapfrogging motion takes place.
In the previous section we have seen how to construct an initial
condition for a single vortex ring. This approach can also be
used to create an initial condition for multiple vortex rings, 
through the addition of the complex variable obtained as the 
stereographic projection of the spin ${\bf n}.$ Explicitly,
${\bf n}$ is a unit vector and the associated point on the
two-sphere can be specified by the Riemann sphere coordinate
\be
\zeta=\frac{n_1+in_2}{1+n_3}.
\label{rsphere}
\ee
Note that the vacuum value ${\bf n}={\bf k}$ maps to the point $\zeta=0$
and the value at the centre of the core 
${\bf n}=-{\bf k}$ maps to the point at infinity $\zeta=\infty.$
Let $\zeta^{(j)},$ for $j=1,\ldots,m,$ denote the Riemann sphere
coordinates obtained from (\ref{rsphere}) by taking the spin ${\bf n}$
to be the initial condition of a single vortex ring with radius
$R^{(j)}$ and position $z_0^{(j)}.$
The set of $m$ single vortex rings are coaxial and 
are taken to have mutual separations that are greater than the core diameter. 
Taking the sum $\zeta=\sum_{j=1}^m\zeta^{(j)}$ and inverting 
(\ref{rsphere}) yields an initial condition for 
$m$ coaxial rings.

\begin{figure*}[h]\begin{center}
\includegraphics[width=16.5cm]{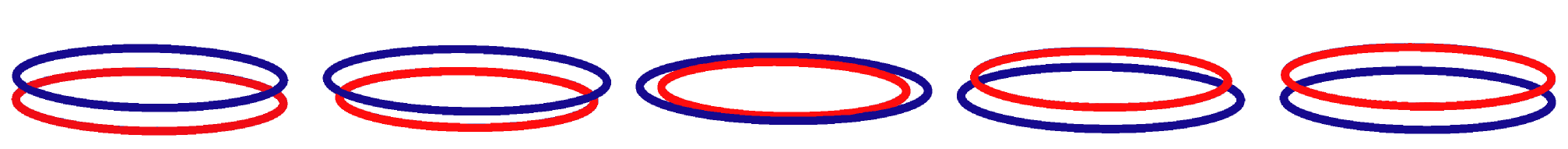}
\caption{
A pair of leapfrogging magnetic vortex rings.
The isosurface $n_3=0$ is displayed at equal time intervals
(from left to right) from 
$t=0$ to $t=1440.$ Initially the two rings have equal radii 
$R=45$ and positions $z_0=\pm 4.$ 
There is an applied electric current $J=0.026$ to freeze 
the centre of mass motion. The colouring of the isosurfaces is
to aid identification of the two rings throughout the 
evolution.
}
\label{fig-leapfrog}\end{center}\end{figure*}

In the first image in Figure~\ref{fig-leapfrog} we
display the initial condition, obtained using the above procedure, 
for two rings with equal radii 
$R=45$ and starting positions $z_0=\pm 4.$
The two thin tubes correspond to
the isosurface $n_3=0,$ identifying the vortex cores. The
two components of this isosurface
are coloured red and blue in order to aid identification of the
two vortex rings throughout the motion.
\begin{figure}[h]\begin{center}
\includegraphics[width=8cm]{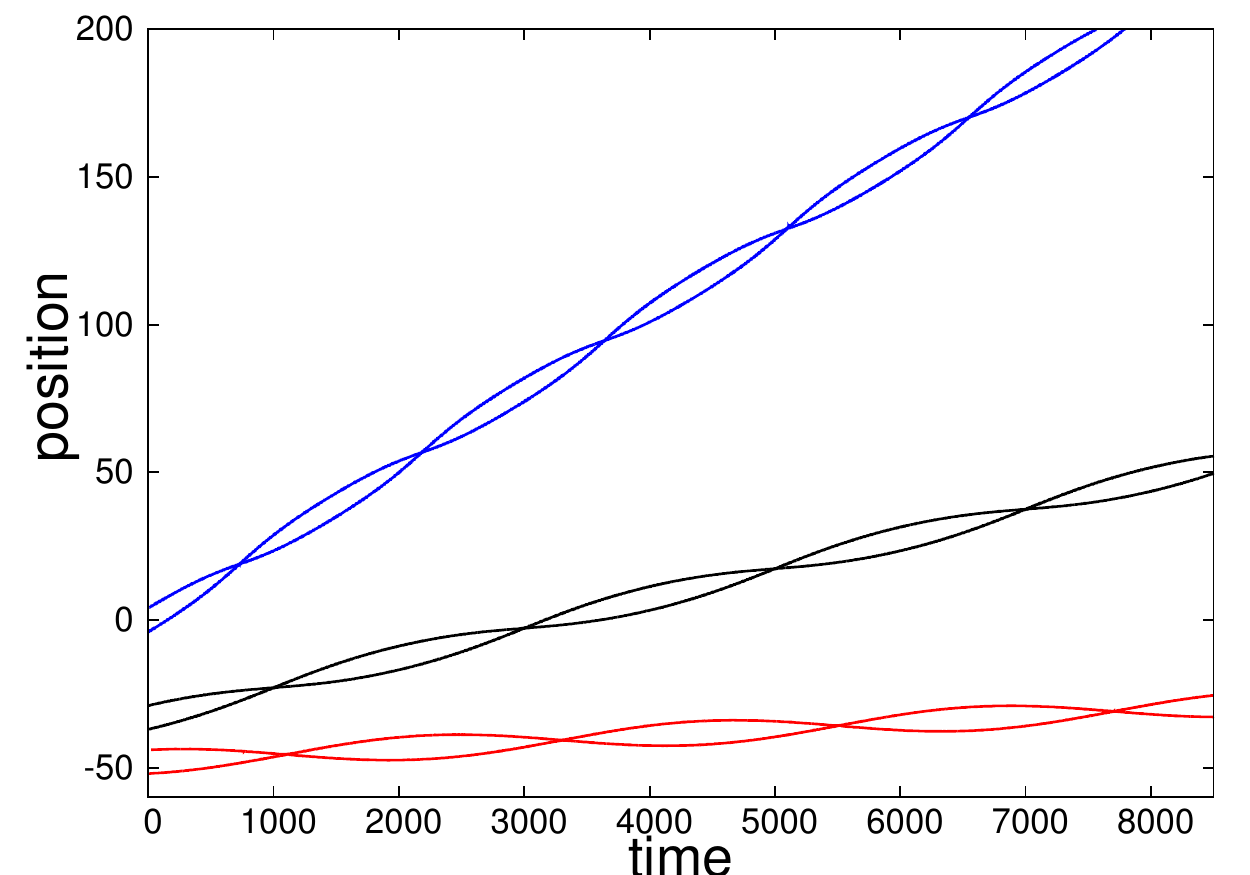}
\caption{
The positions of pairs of leapfrogging vortex rings in the absence of
an electric current.
The graph shows the positions along the $z$-axis as a function of time 
for pairs of leapfrogging vortex rings with initial equal radii $R$
and an initial separation in the $z$ direction equal to 8.
Upper blue curves have $R=45$, 
middle black curves have $R=115$ and 
lower red curves have $R=523$.
}
\label{fig-position}\end{center}\end{figure}
The images in Figure~\ref{fig-leapfrog} show the
time evolution, obtained by numerical solution of (\ref{llj}) 
in the presence of an electric current $J=0.026,$ 
applied to freeze the centre of mass motion.
These images reveal that the bottom (red) ring shrinks and speeds up 
until it 
passes through the top (blue) ring, after which it expands so that the initial
configuration is recovered but with an exchange of the two rings.
This is the famous leapfrogging motion discussed by Helmholtz \cite{He}
in the mid-nineteenth century, in the context of vortex rings in fluids.
Here we have shown, for the first time, 
that the same phenomenon takes place in the nanoscopic 
world of ferromagnetic spin structures, as modelled by the Landau-Lifshitz
equation. In the absence of dissipation, perpetual leapfrogging is 
expected. Our numerical results support this expectation, with 
our longest simulations producing around a dozen leapfrogging events
with no discernible deviation from a periodic motion. 
\begin{figure}[ht]\begin{center}
\includegraphics[width=16.5cm]{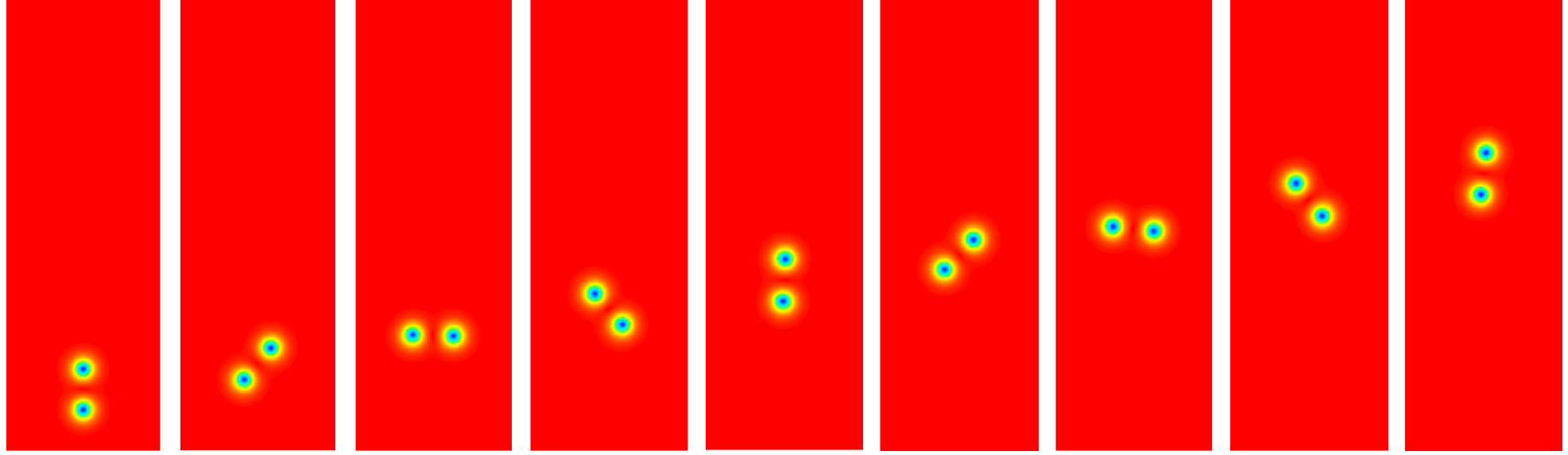}
\caption{
A cross-section in the $(\rho,z)$ plane showing
plots of $n_3,$  at equal time intervals (from left to right) 
from $t=0$ to $t=4000,$
with colour scale from blue to red corresponding 
to values from -1 to 1.
In each plot the displayed region is $(\rho,z)\in[100,130]\times[-45,45].$
The cross-section contains a pair of Skyrmions that rotate around each other 
and propagate in the $z$-direction. The associated three-dimensional vortex 
rings are obtained by rotating these results about the $z$-axis. 
From the three-dimensional perspective each ring has an initial radius $R=115$
and the rings leapfrog each other.
}
\label{fig-xsection}\end{center}\end{figure}
\begin{figure*}[ht]\begin{center}
\includegraphics[width=8.1cm]{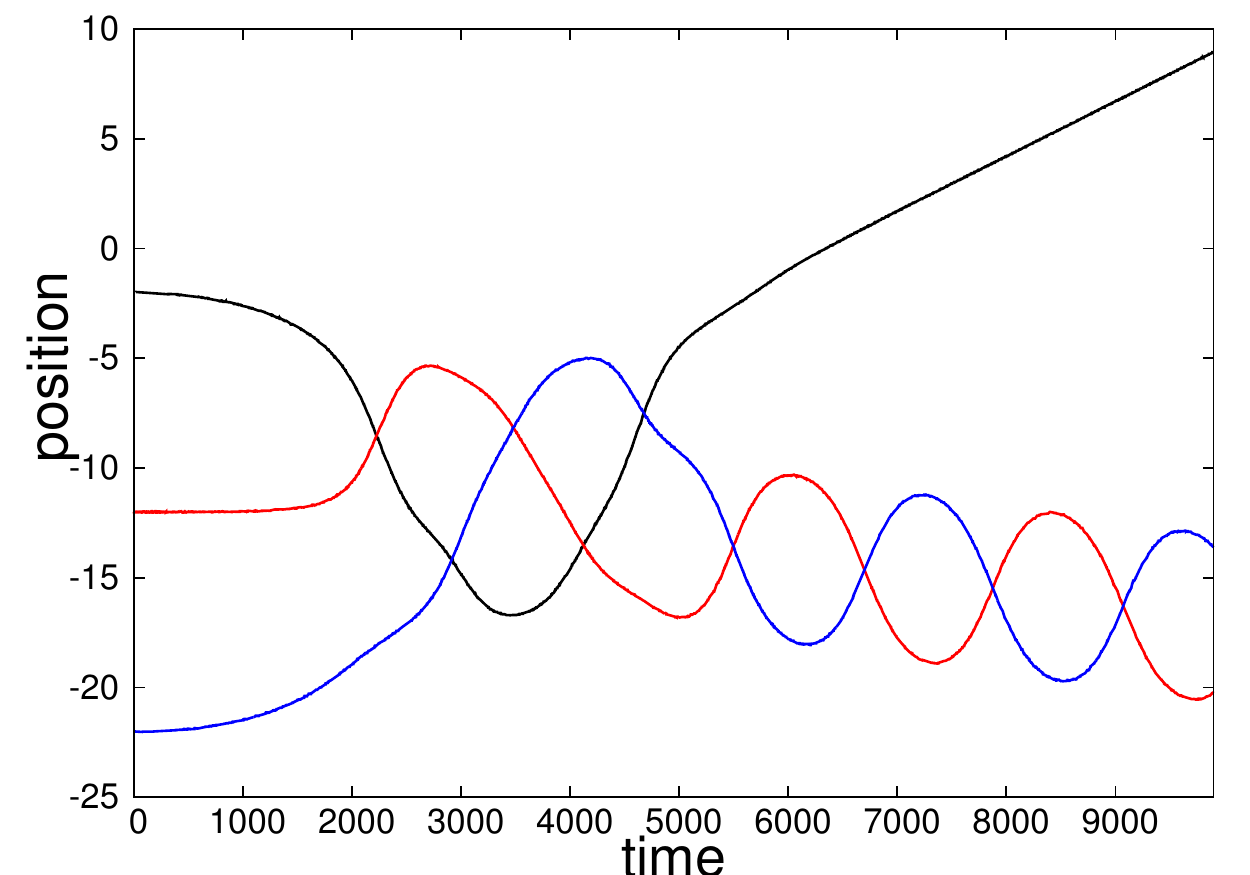}
\includegraphics[width=8.1cm]{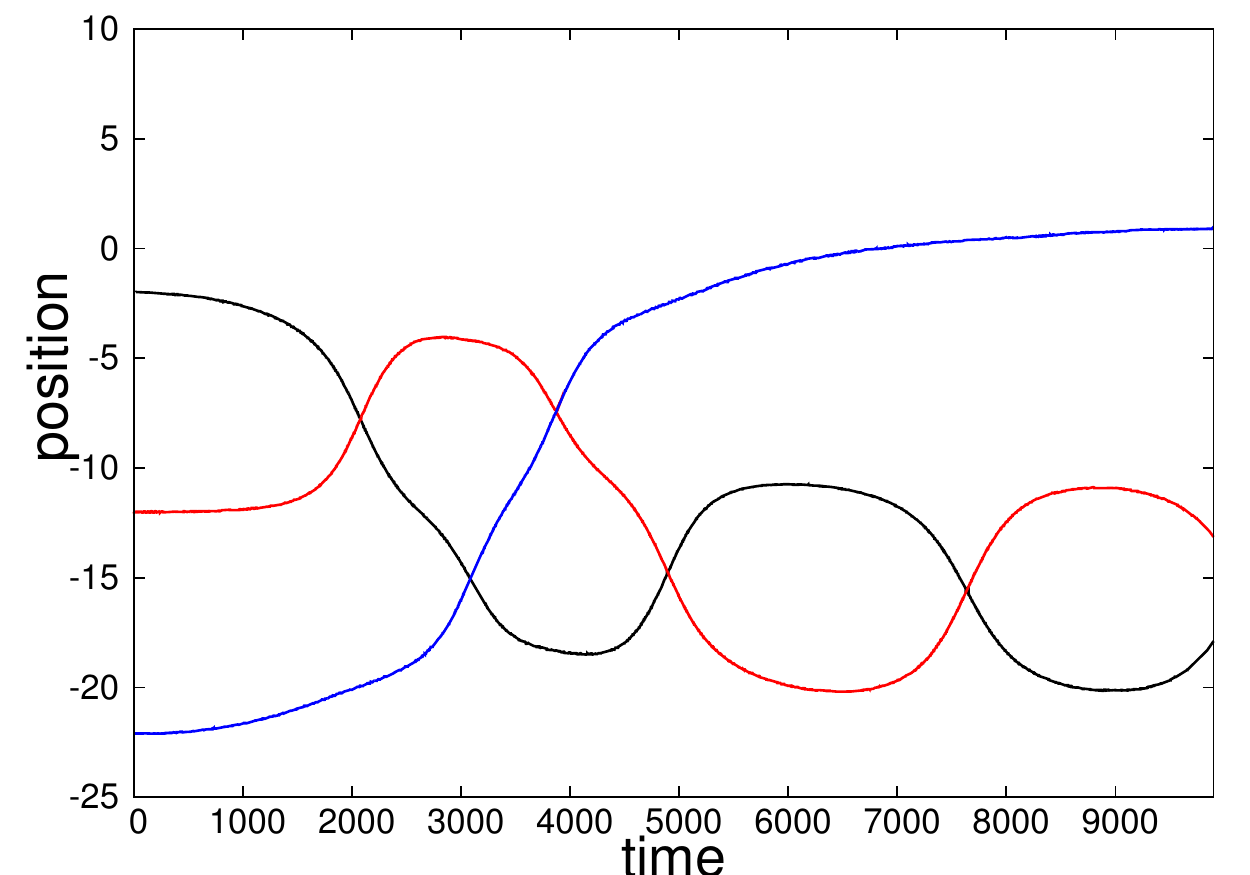}
\caption{
Trios of leapfrogging vortex rings in the presence
of an electric current $J=0.026,$ to freeze the centre of mass
motion.
The positions along the $z$-axis as a function of time 
for three vortex rings with initial equal radii $R=45.$
In the left image the rings are equally spaced with initial positions 
$z_0=-2,-12,-22.$ 
In the right image the initial separation between the two lowest
rings has been increased by $1\%$ to give the initial positions 
$z_0=-2,-12,-22.1$ 
The drastic difference in the resulting evolution demonstrates
the sensitivity to the initial conditions expected of chaotic dynamics.
}
\label{fig-trio}\end{center}\end{figure*}

\begin{figure}[ht]\begin{center}
\includegraphics[width=5cm]{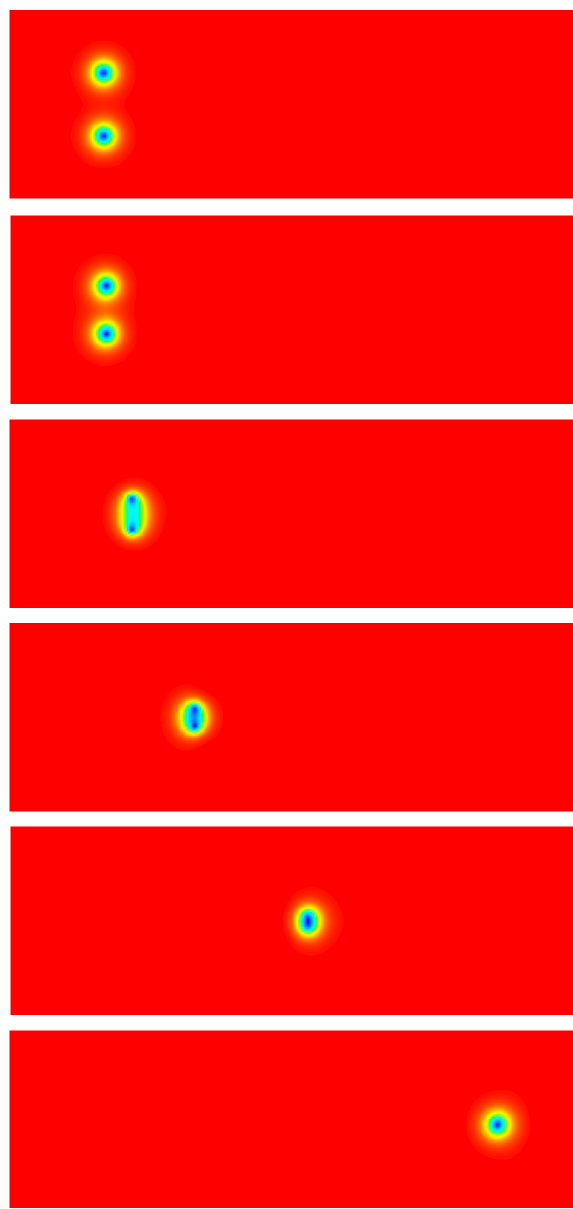}
\caption{
The cross-section for a head-on collision of two vortex rings, 
with initially equal radii $R=45$ and opposite orientations.
The three-dimensional vortex rings are obtained by rotating these
cross-sections around the $z$-axis.
The initial axis positions are $z_0=\pm 5.$
The plots show $n_3,$ 
at equal time intervals (from top to bottom) from $t=0$ to $t=228,$
with colour scale from blue to red corresponding 
to values from -1 to 1.
In each plot the displayed region is $(\rho,z)\in[30,120]\times[-15,15].$
The head-on collision generates a rapidly expanding single ring.
}
\label{fig-headon}\end{center}\end{figure}
A Galilean transformation can be applied to convert the above result to
the situation in which there is no applied electric current, so that
the leapfrogging pair propagate up the $z$-axis. 
This yields the upper blue curves in Figure~\ref{fig-position} for the 
$z$-positions of the pair of vortex rings.
These curves display multiple leapfrogging events, identified by
the crossing points of the two curves.  
Increasing the radius of the vortex rings to $R=115$ produces the 
middle black curves in Figure~\ref{fig-position}.
The associated effective two-dimensional simulation 
(obtained as the half-plane $(\rho,z)$ by taking a cross-section)
is presented in Figure~\ref{fig-xsection}, for the case of 
no electric current.
In the cross-section, 
this shows a pair of Skyrmions rotating around each other whilst
drifting in the positive $z$-direction.
The three-dimensional vortex rings are obtained by 
rotating these cross-sectional results about the $z$-axis. 
Under this identification, a pair of cross-sectional Skyrmions that
rotate around each other maps to a leapfrogging event.
 
A further increase of the radius 
to $R=523$ produces the lower red curves
in Figure~\ref{fig-position}. 
This set of results demonstrates that increasing the
ring radius has a significant influence on the propagation speed of the
pair, which decreases with increasing ring radius.
However, there is much less of an impact
on the leapfrogging period, which is mainly determined by the separation
of the rings, and increases with this separation.  
The results in \cite{Kom} suggest that
the frequency of the leapfrogging motion should vary with the 
inverse of the square of the separation between the rings, 
and it may be possible to 
derive such a relation analytically using the methods described there.

The similarity between magnetic vortex rings and those in fluids extends to 
other aspects beyond leapfrogging. Below we shall demonstrate some examples of
this correspondence. 
 
By applying a thin-cored approximation,  
studies \cite{Ko} of three or more coaxial fluid vortex rings 
have shown that there are regimes of chaotic behaviour in which the
evolution is very sensitive to the initial condition.
We now investigate the same issue for magnetic vortex rings, but using the
full nonlinear partial differential equations, 
rather than a thin-cored approximation.
The left image in Figure~\ref{fig-trio} displays the evolution 
of the positions along the $z$-axis of a trio of
equal radii $R=45$ vortex rings that initially are equally spaced.
There is an applied electric current $J=0.026,$ to keep all three
rings inside the simulation region for a reasonable length of time.
It can be seen that after some mutual leapfrogging, the asymptotic state
contains a pair of leapfrogging rings plus a decoupled single ring,
which is the one that was initially at the top.

The right image in Figure~\ref{fig-trio} displays the resulting evolution 
following a tiny change in the initial condition to increase the 
initial separation of the two lowest rings by $1\%.$ 
It can be seen that this tiny change results in a drastic difference
in the dynamics. There are fewer mutual leapfrogging events and this time
the single ring that decouples is the ring that was initially at the
bottom not the top. The period of the asymptotic leapfrogging pair has
also increased to about twice the value found in the first case,
indicating that this pair has an increased separation. 
This sensitivity to the initial conditions is in line with the results obtained
for fluid vortex rings \cite{Ko} 
and is an indication of the presence of chaotic dynamics. 
The fact that we
are able to reproduce this behaviour further strengthens the analogy
between magnetic vortex rings and their fluid counterparts. 

We have also studied the head-on collision of two coaxial magnetic vortex 
rings that are initially moving in opposite directions and have
equal radii. 
A vortex ring moving along the negative $z$-axis is obtained by replacing
the initial $q=1$ Skyrmion cross-section by an anti-Skyrmion with $q=-1.$
In Figure~\ref{fig-headon} we display the initial cross-section, with
axis positions $z_0=\pm 5$ and a common radius $R=45$, together
with the subsequent evolution of the cross-section.
We find that the two rings merge into a rapidly expanding and thinning ring,
as seen in the initial stages of experiments on fluids \cite{LN}.

In terms of the cross-section, the single structure that forms from
the collision has Skyrmion number $q=0,$ as it is formed from the merger 
of a Skyrmion and an anti-Skyrmion.
This can be understood within the two-dimensional system 
as the formation of a non-topological soliton, which is an object
that has been studied in \cite{PZ}. 
The non-topological soliton is a stationary solution that
has the form (\ref{ansatz}) but with $q=0.$ The resulting ordinary
differential equation for the profile function, (\ref{ode}) with $q=0,$
is now subject to the boundary condition $f'(0)=0,$ rather than
$f(0)=\pi,$ which was required for $q\ne 0.$
The fact that the vortex ring with a $q=0$ cross-section has an expanding
radius can be understood from the above description of its formation.
Namely, a perturbation of the $q=0$ cross-section is a Skyrmion anti-Skyrmion
pair in cross-section and one expects the cross-sectional dynamics to be
qualitatively similar to the two-dimensional dynamics of a planar
Skyrmion anti-Skyrmion pair. Such a pair generates a common translational 
motion, in the same manner as a vortex anti-vortex pair in fluid dynamics,
which turns into an expanding ring when considered as a cross-section that
is to be rotated around the $z$-axis to provide the full three-dimensional
configuration. 

The head-on collision of vortex rings provides another demonstration
of the similarity between vortex rings in fluids and their magnetic analogues.
It would be interesting to extend our simulations beyond the axially
symmetric regime to see if the non-axial instability and subsequent 
production of small rings via reconnection, seen in the latter stages of 
the fluid experiments \cite{LN}, can be reproduced in the
Landau-Lifshitz equation.

\section{Vortex rings in the Landau-Lifshitz-Gilbert equation}\news\label{sec-LLG}
So far we have neglected dissipation, but this can be included in the theory 
by extending our simulations to the Landau-Lifshitz-Gilbert equation.
In the absence of an electric current, this equation reads \cite{TKS}
\be
\frac{\partial {\bf n}}{\partial t}={\bf n} \times 
\{\nabla^2{\bf n} + A({\bf n}\cdot{\bf k}){\bf k}\}
-\lambda {\bf n}\times({\bf n}\times
\{\nabla^2{\bf n} + A({\bf n}\cdot{\bf k}){\bf k}\})
\label{llg}
\ee
where $\lambda>0$ is the damping constant.

\begin{figure}[ht]\begin{center}
\includegraphics[width=10cm]{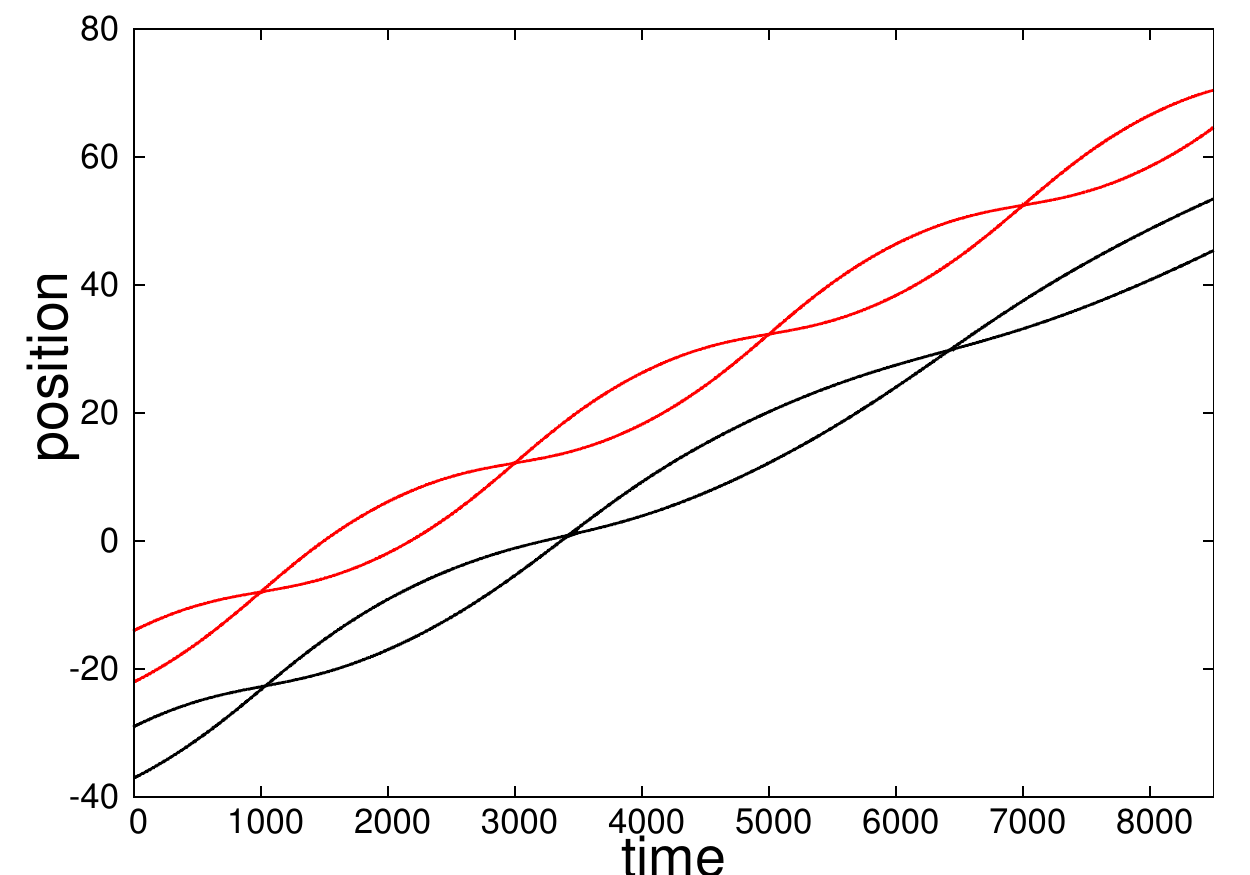}
\caption{The upper red curves show the propagation of a leapfrogging pair of
vortex rings without dissipation. The lower black curves show the result
for the same pair after the inclusion of a small damping constant:
to aid visualization, the lower black curves have been shifted down.
The damping has very little effect on the radii of the vortex rings
but their thickness decreases, which results in a continual
increase in the period of leapfrogging.
}
\label{fig-llg}\end{center}\end{figure}

Once damping is included,
 the perpetual leapfrogging of a pair of vortex rings is
replaced by a finite number of leapfrogs, with this number decreasing
as the damping constant is increased. 
Figure~\ref{fig-llg} provides an illustrative example of this phenomenon
for two vortex rings with initial radii $R=115$ and a separation of 8 along
the $z$-axis. The upper red curves are the position curves for the 
two vortex rings in the absence of dissipation, that is, 
solutions of (\ref{llg}) with $\lambda=0.$
The lower black curves display the positions of the vortex rings
using identical initial conditions, but this time for solutions of 
(\ref{llg}) with the small damping constant
$\lambda=0.0001.$ To aid visualization, this second pair of curves have
been shifted down to avoid any overlap with the first set of curves.

Damping does not have a significant influence on the radius of a vortex ring 
but instead the main effect is to reduce the thickness of the vortex ring.
In terms of the planar cross-section, this is a reduction of the core size
of the Skyrmion. For a leapfrogging event, this reduction in core size 
means that the distance between the two Skyrmions has increased relative
to the core size, which increases the period of the rotating pair. 
This phenomenon is clearly demonstrated in Figure~\ref{fig-llg},
where it is evident that the dissipation produces a continual increase in the 
leapfrogging period. 
Again this has a direct
analogy in fluid dynamics, where it has been observed \cite{RS} 
that the number of leapfrogs increases with Reynolds number.
Perpetual leapfrogging of vortex rings in fluids can only
occur in inviscid flow, which in our magnetic analogy corresponds to the 
Landau-Lifshitz equation without dissipation.

In materials that are currently used in experiments in standard conditions,
the damping constant can be as large as $\lambda=0.02.$ 
For a frequency of precession $\omega$ the effect of damping is that
the precession spirals down on a time scale of order $(\lambda\omega)^{-1}.$
Given this relation and our above results for small damping it 
will therefore be a considerable challenge to create experimental
conditions under which leapfrogging has time to take place.

\section{Conclusion}\news\label{sec-con}
In this paper we have presented the results of numerical simulations of
vortex ring dynamics in the Landau-Lifshitz equation, describing the
evolution of the local magnetization in a ferromagnetic continuum.
We have observed many similarities between the dynamics of magnetic vortex
rings and those in fluids, including the famous leapfrogging locomotion
of a pair of vortex rings. 

Although it is beyond the present paper, it would be interesting to
study how our results are modified by additional effects 
that are can be included within the Landau-Lifshitz equation to improve its
description of physical materials.
An example is the dipole-dipole 
interaction: though as a non-local effect, this would require a substantial 
increase in computation and thus constitutes a future project. 
Another example is a realistic modelling of temperature dependence. 
However, as long as the Landau-Lifshitz model remains a valid mean field 
description, temperature effects can be accounted for via a 
renormalization of the parameters of the theory. 
As we have observed, the qualitative features of vortex ring 
dynamics appear to be universal, with many similarities between our results and
known results in fluid dynamics.
Hence we expect the qualitative features to remain robust to the inclusion
of any additional terms, 
although there may be some 
quantitative differences that could be important in 
attempting any experimental observations of these phenomena.

There are enormous technical challenges that must be overcome before
the experimental creation and observation of magnetic vortex rings
can be achieved, as this requires the manipulation of highly nonlinear
excitations involving a large number of spin-flips.
As we have discussed, the cross-section of a magnetic vortex ring has 
the non-trivial topology of a Skyrmion, so the recent experimental
observations of Skyrmions in chiral magnets \cite{Yu},
 with a size of the order of $100$ nanometres, offers some
encouragement that the complex phenomena predicted here might be 
observed in future experiments.

Current-driven domain wall motion is an important phenomenon in the
rapidly developing field of spintronics, with applications to high density
memory storage. Vortex rings provide a complementary companion to domain
walls in that a single vortex ring is in uniform motion without a current,
but can be brought to rest by the application of a current. 
Magnetic vortex rings may therefore have a future role to play within
spintronic devices. 

There are also opportunities to exploit the 
interactions of multiple vortex rings.   
As mentioned earlier, it has been shown \cite{SM} 
that within the thin-cored approximation of 
fluid vortex rings, the periodic leapfrogging motion of a pair of vortex 
rings generates a geometric phase. 
It is expected that such results can be extended to
more sophisticated vortex ring interactions, such as those considered here.
This is related to the existence of exotic exchange statistics
for leapfrogging motion \cite{Ni}, that follows from the braiding of
vortex loops, to yield a three-dimensional version of the phenomenon
that produces anyon statistics in two dimensions.
The relevant mathematical generalization of the braid group is the
McCool group \cite{Mc,Go} and leapfrogging magnetic vortex rings
provide a physical realization of this group.
Given the importance of anyons in applications to topological quantum
computing, we may speculate on similar potential applications for 
leapfrogging magnetic vortex rings.
As we have observed, a pair of leapfrogging rings can be held in position by 
a spin-polarized electric current, allowing exotic exchange statistics 
to be physically realised. By turning off the current the vortex rings are 
released, allowing the result to be read.\\    

Finally, we make some comments on the important differences between vortex 
rings in the Landau-Lifshitz equation and related structures, 
known as Hopf solitons \cite{MS}, 
that exist in relativistic theories of a  
three-component unit vector.

Consider any theory, in three-dimensional space, defined for a 
three-component unit vector ${\bf n},$ with the boundary condition
${\bf n}\rightarrow {\bf k}$ as $|{\bf x}|\rightarrow \infty.$
Irrespective of the equations of motion, such a theory has a 
conserved integer-valued topological charge $Q,$ known as the Hopf invariant.
Physically, $Q$ is the linking number of two curves obtained as the preimages
of any two distinct generic values of ${\bf n}.$

The concept of the Hopf invariant is applicable to vortex
ring solutions of the Landau-Lifshitz equation, but for 
all the vortex rings studied in this paper the Hopf invariant is zero.
In fact $Q=0$ follows directly from the restriction to axial symmetry.
Note that there is some potential for confusion here, 
particularly in relation to Hopf solitons, 
because it is common to use the term axial symmetry when the intended
meaning is actually equivariance, rather than symmetry.
Such equivariance means that a rotation of the cylindrical angle $\theta$
through an angle $\alpha$ can be compensated by an internal rotation of the 
two-component vector $(n_1,n_2)$ through an angle $m\alpha,$ for some
non-zero integer $m.$ Strict axial symmetry, as used in the current paper, 
corresponds to the case $m=0,$ so that no compensating internal rotation
is needed. 

In terms of our earlier description of a vortex ring, as
a planar Skyrmion rotated around the $z$-axis, then a vortex ring with Hopf 
invariant $Q$ is obtained if the $(n_1,n_2)$ components of the planar 
Skyrmion are rotated through an angle $2\pi Q$ as the planar Skyrmion
is rotated through an angle $2\pi$ around the $z$-axis. 
It is possible to study Landau-Lifshitz vortex rings with a
non-zero Hopf invariant \cite{Co,Su}, but the properties of a single ring
are very similar to rings with $Q=0,$ 
so the Hopf invariant is not an important ingredient 
in this setting. This contrasts markedly with the case of relativistic Hopf
solitons, where the Hopf invariant plays a vital role, and vortex ring 
structures exist only if $Q\ne 0.$
Another important difference in relativistic theories is that Hopf solitons
can be static. A moving Hopf soliton has an arbitrary speed (less than
the speed of light) and is simply obtained from the static solution by
performing a Lorentz boost.  

The most detailed studies of Hopf solitons have been performed in 
the Skyrme-Faddeev model \cite{FN}, where solitons with $Q=1$ and $Q=2$ are 
axially equivariant (as described above) and have a vortex ring structure.
The size of these vortex rings is not arbitrary, but is fixed by the 
parameters in the theory. The ratio of the radius of a vortex ring to its
thickness is independent of these parameters, and so there is no 
scope for a large thin ring, of the kind studied in the present paper.
Moreover, 
Hopf solitons form bound states, with the energy of the static $Q=2$ 
soliton being less than twice the energy of the static $Q=1$ soliton.
For larger values of $Q$ Hopf solitons form knotted and linked configurations
\cite{BS,HS}, rather than axial vortex rings.
This attraction between Hopf solitons has important consequences for the 
relativistic dynamics of these vortex rings, 
as studied recently in \cite{HPJP}. These simulations reveal that
 vortex rings tend to merge, or bounce off each other, depending upon 
internal orientations and impact parameters, but they do not leapfrog. The 
first order dynamics of the Landau-Lifshitz equation, discussed in the current
paper is therefore very different to the relativistic dynamics presented
in \cite{HPJP}, even though, at first glance, there may appear to be some 
similarities between the two systems.  

\section*{Acknowledgements}
\noindent 
AJN is supported by a CNRS PEPS collaboration grant, by a Region
Centre Recherche d'Initiative Academique grant, by a
Chinese-French Scientific Exchange program Cai Yuanpei, and
by a research grant through the Qian Ren Program at BIT.

\noindent PMS thanks Fred Cohen for useful discussions,
and is grateful for the hospitality of the KITP in Santa Barbara.

\end{document}